\documentclass[12pt]{iopart}

\usepackage{graphicx}
\usepackage{epsfig}
\usepackage{color}
\usepackage{subfig}

\begin{document}

\title{Markovian master equation for nonlinear systems}
\author{O de los Santos-S\'anchez$^{(1)}$, J R\'ecamier$^{(2)}$ and R J\'auregui$^{(1)}$} 

\address{$^{(1)}$ Instituto de F\'{\i}sica, Universidad Nacional Aut\'onoma de M\'exico, Apdo. Postal 70-542, 04510 M\'exico, Distrito Federal, M\'exico}
\address{$^{(2)}$ Instituto de Ciencias F\'{\i}sicas, Universidad Nacional Aut\'onoma de M\'exico, Apdo. Postal 48-3, Cuernavaca, Morelos 62251, M\'exico}

\eads{\mailto{octavio.desantos@gmail.com}, \mailto{pepe@fis.unam.mx}, \mailto{rocio@fisica.unam.mx}}

\begin{abstract}
Within the f-deformed oscillator formalism, we derive a Markovian master equation for the description of the damped dynamics of nonlinear systems that interact with their environment. The applicability of this treatment to the particular case of a Morse-like oscillator interacting with a thermal field is illustrated, and the decay of quantum coherence in such a system is analyzed in terms of the evolution on phase space of its nonlinear coherent states via the Wigner function.
\end{abstract}

\maketitle

\section{Introduction}

In recent times, the notion of nonlinear quantum deformed systems and their associated algebras has played a preponderant role in many active fields of quantum physics, ranging from atomic to molecular physics, and particularly in quantum optics. This tendency started with the idea of deforming the usual commutation relation of the harmonic oscillator by means of what is called a q-deformation, from which, of course, the concept of q-deformed oscillators emerges \cite{sklyanin, Bied}. More recently, $e.$ $g.$, a new algebraic formalism for supersymmetric and shape-invariant systems was proposed in \cite{aleixo} to construct nonlinear coherent states for these kind of systems. Indeed, these ideas were conceived from the necessity to deal with additional physical effects encountered in generalized quantum systems other than the harmonic oscillator. 

A novel nonlinear algebraic theory, which will be of interest to us, concerns a special quantum deformation process on the harmonic oscillator algebra, namely, the so-called {\it f-deformed oscillator formalism} introduced by Man'ko and coworkers \cite{manko1,manko2}. Within this algebraic scheme, which contains the q-deformed algebra as a particular case, the knowledge of the respective deformation function that fixes the nonlinearity of the system under study allows for the dynamical and statistical analysis of it in terms of its deformed (nonlinear) coherent states in a straightforward way. The wide applicability of this nonlinear algebraic formalism has been demonstrated in a manifold of circumstances. Among the most representative, we can quote the paper of Man'ko \etal \cite{manko3} about the Weyl-Wigner-Moyal representation for f-oscillators, in which the well-known Kerr-like nonlinearity of media was taken into account; the description of the center-of-mass motion of a laser-driven trapped ion \cite{matos, manko4, kis}; generalizations of the Jaynes Cummings model in which the interaction between a two or three-level atom and the radiation field is nonlinear in the field variables \cite{crnugelj1,crnugelj2}, as well as including Kerr type nonlinearities \cite{octa1,faghihi}; the relation between the deformation function of the f-deformed oscillator and the two-dimensional harmonic oscillator on the flat space and on the sphere \cite{mahdifar}; and generalizations of coherent states for confining systems such as the symmetric  P\"oschl-Teller potentials \cite{bagheri,octa2} and the Morse potential \cite{recamier1,recamier2,recamier3,recamier4,octa3,octa4}.\\

 Another issue that has attracted much attention in recent years is the phenomenon of quantum decoherence. As known, since a given physical system cannot be completely isolated from its environment, a more realistic description of it has to take into consideration the effects of dissipation due to such an interaction. There have been some interesting discussions on the decay of quantum coherence of deformed nonlinear oscillators for several types of interaction between the system and its environment in the frame of a master-equation treatment. For instance, let us mention the work of Mancini \cite{mancini} where a general master equation describing the damped dynamics of an f-deformed oscillator was derived by considering a linear interaction between the system and its environment; the latter being a heat bath. Along this line, Isar \etal \cite{isar1,isar2} have obtained, on the one hand, a Lindblad-like master equation \cite{lindblad} for the standard harmonic oscillator interacting with an environment via a deformed interaction, and, on the other hand, a Markovian master equation for describing a f-deformed oscillator coupled to a thermal bath with a nonlinear coordinate-momentum coupling.
 
Motivated by the aforementioned studies, we put forward an alternative way of describing the damped dynamics of a nonlinear system, and that is to build the respective generalized Markovian master equation on the basis of a nonlinear dipolar-type coupling between the system and its environment. Here, besides incorporating the archetypical influence of the deformation function, the strength of the coupling also involves, as it should, an explicit dependence on the dipole transition probability through the corresponding matrix element for given particular systems.
We work out explicitly the case of an environment that can be viewed as a thermal reservoir. \\

The organization of this paper is the following: In section 2, a brief description of the basics of the f-deformed oscillator formalism is reminded. In section 3, we introduce a Hamiltonian model for the description of a dipolar-type interaction between a deformed oscillator (for any deformation function with physical meaning) and its environment; the latter is considered to be composed by an infinite set of harmonic oscillators. In section 4, a generalized master equation governing the evolution of such an oscillator as an open quantum system is derived within the context of the Born and Markov approximations. As a particular example, in section 5 the master equation thus obtained is applied to describe the damped dynamics on phase space of the so-called Morse-like oscillator coupled to a thermal bath, using the pictorial representation given by the Wigner quasi-distribution function.  And finally, in section 6 some conclusions are given.

\section{The f-deformed oscillator}

According to Man'ko \etal  \cite{manko1,manko2}, an f-deformed oscillator is a nonlinear quantum system modeled by a Hamiltonian of the harmonic oscillator form
\begin{equation}
\hat{H}_{S} = \frac{\hbar \Omega_{0}}{2}(\hat{A}^{\dagger}\hat{A}+\hat{A}\hat{A}^{\dagger}),
\label{eq:defham1}
\end{equation}
where the deformed boson annihilation and creation operators $\hat{A}$ and $\hat{A}^{\dagger}$, respectively, are defined by deforming the standard harmonic operators $\hat{a}$ and $\hat{a}^{\dagger}$ through the non-canonical transformation
\begin{equation}
\hat{A}=\hat{a} f(\hat{n}) = f(\hat{n}+1)\hat{a}, \qquad \hat{A}^{\dagger} = f(\hat{n}) \hat{a}^{\dagger} = \hat{a}^{\dagger} f(\hat{n}+1),
\label{eq:defops}
\end{equation}
where, in turn, $\hat{n}=\hat{a}^{\dagger}\hat{a}$ is the usual number operator, and the operator function $f(\hat{n})$, which is assumed to be real, is a deformation function depending on the level of excitation of the system.

Substitution of definition (\ref{eq:defops}) into the deformed Hamiltonian (\ref{eq:defham1}), together with the use of the well-known canonical commutation relation $[\hat{a},\hat{a}^{\dagger}]=1$, gives us an equivalent expression for the said Hamiltonian written in terms of the number operator only, i.e., 
\begin{equation}
\hat{H}_{S} = \frac{\hbar \Omega_{0}}{2}\left( (\hat{n}+1)f^{2}(\hat{n}+1)+\hat{n}f^{2}(\hat{n}) \right),
\label{eq:defham2}
\end{equation}
which, needless to say, is diagonal on the basis of number-states. Similarly it is also found that the set of operators $\{ \hat{A},\hat{A}^{\dagger}, \hat{n} \}$ obeys the commutation relations
\begin{equation}
\left[ \hat{A}^{\dagger},\hat{n} \right] = -\hat{A}^{\dagger}, \qquad \left[ \hat{A},\hat{n} \right] = \hat{A},
\end{equation}
and
\begin{equation}
\left[ \hat{A},\hat{A}^{\dagger} \right] = (\hat{n}+1)f^{2}(\hat{n}+1)-\hat{n}f^{2}(\hat{n}).
\end{equation}
We can see that, unlike the harmonic oscillator case, the Hamiltonian model (\ref{eq:defham2}) of an f-deformed oscillator will no longer be, in general, a linear function of the number operator, since additional powers of it may take place; this is the reason why such systems are usually referred to as nonlinear. On the other hand, it is also clear that the commutator between $\hat{A}$ and $\hat{A}^{\dagger}$ is no longer a c-number, but it may become a rather complicated function of the number operator, depending naturally on the particular choice of the deformation function $f(\hat{n})$. In the limit $f(\hat{n}) \to 1$ this f-deformed algebraic scheme contracts to the usual harmonic oscillator algebra.\\

It is also important to point out that the deformed operators have the following effect on the number operator basis $|n\rangle$:
\begin{eqnarray}
\hat{A} |n\rangle & = & f(n)\sqrt{n} |n-1\rangle, \label{equ:def1} \\
\hat{A}^{\dagger} |n\rangle & = & f(n+1) \sqrt{(n+1)} |n+1\rangle, \label{equ:def2}
\end{eqnarray}
thus changing the number of quanta in $\pm 1$, and their corresponding matrix elements are modified through the deformation function $f$. That is, they create and annihilate a single anharmonic quantum within the context of the nonlinear system described by Hamiltonian (\ref{eq:defham1}). This is also highlighted by the commutation property
\begin{equation}
\left [\hat{H}_{S}, \hat{A} \right]   =   -\hbar \Omega(\hat{n})\hat{A},  \quad \left [\hat{H}_{S}, \hat{A}^{\dagger} \right]   =  \hbar \hat{A}^{\dagger}\Omega(\hat{n}),
\end{equation}
where the operator
\begin{equation}
\Omega(\hat{n}) = \frac{\Omega_{0}}{2} \left((\hat{n}+2)f^{2}(\hat{n}+2)-\hat{n}f^{2}(\hat{n}) \right)
\label{equ:freqn}
\end{equation}
has the physical meaning of being the frequency separation between adjacent energy states, as inferred from Hamiltonian (\ref{eq:defham2}); as a matter of fact, by definition, any set of operators satisfying the above property is said to be the set of eigenoperators corresponding to the Hamiltonian of the specific quantum system of interest \cite{gardiner}. Indeed, the deformed operators can be understood as jump operators that would allow us to describe anharmonic transitions in the frame of nonlinear systems. \\

In what follows we shall make use of the present formalism for the description of the nonlinear nature of systems whose discrete energy spectrum is not equally spaced. Its utility will be illustrated both in establishing a Hamiltonian model describing the interaction of such systems with an external environment and in putting forward the corresponding equation of motion so as to examine their temporal evolution based on the density operator method.

\section{Oscillator-environment interaction Hamiltonian}

Let a nonlinear system interacting with a reservoir of harmonic oscillators be described by the Hamiltonian 
\begin{eqnarray}
\hat{H}= \hat{H}_{S}+\hat{H}_{R}+\hat{H}_{Int}.
\label{equ:totalham}
\end{eqnarray}
Here, $\hat{H}_{S}$ is the Hamiltonian of the f-deformed oscillator given by Eq.~(\ref{eq:defham2}), and $\hat{H}_{R}$ is the reservoir Hamiltonian; to be more specific, the later is considered to represent a radiation field composed by an infinite set of harmonic oscillators which takes on the role of the environment. In the following, this will be modeled by the Hamiltonian
\begin{equation}
\hat{H}_{R} = \hbar \sum_{k} \omega_{k} \hat{b}^{\dagger}_{k}\hat{b}_{k},
\label{equ:reservoir}
\end{equation}
where $\hat{b}^{\dagger}_{k}$, $\hat{b}_{k}$, respectively, are taken to be creation and annihilation operators of such a system of oscillators of frequency $\omega_{k}$ (here the subscript $k$ is a shorthand notation for each mode of radiation). And $\hat{H}_{Int}$ in (\ref{equ:totalham}), referred to as the interaction Hamiltonian, represents the influence of the environment on the system in question. The explicit form of it is proposed to be
\begin{equation}
\hat{H}_{Int} = {\bf d} \cdot {\bf E},
\label{equ:intham}
\end{equation}
which is regarded as an electric-dipole interaction, where ${\bf d}=\gamma_{0} {\bf r}$ is the dipole moment operator associated with the single deformed oscillator (we have chosen ${\bf r}=r\hat{e}_{r}$ to have the direction of the dipole's oscillations along the unit vector $\hat{e}_{r}$), with $\alpha_{0}$ being a physical parameter whose value can be adjusted depending on the physical system under study. The electric field operator ${\bf E}$, on the other hand, is evaluated in the dipole approximation at the position of the oscillator, say ${\bf r}_{0}$. That is, in the Schr\"odinger picture,  
\begin{equation}
{\bf E} = \sum_{ k}\hat{\epsilon}_{ k}\mathcal{E}_{ k}(\hat{b}_{k}e^{i {\bf k \cdot {\bf r}_{0}}}+\hat{b}^{\dagger}_{ k}e^{-i {\bf k}\cdot {\bf r}_{0}}),
\end{equation}
where $\hat{\epsilon}_{k}$ is a unit polarization vector and $\mathcal{E}_{k}=(\hbar \omega_{k}/2\epsilon_{0} V)^{1/2}$, with $V$ being the size of an arbitrary quantization volume.

At this stage, it is convenient to express the magnitude of the dipole moment in terms of the number operator basis as follows:
\begin{equation}
\gamma_{0} r =  \sum_{n,m}|m\rangle \langle m|r|n\rangle \langle n|,
\label{equ:dipole1}
\end{equation}
where the matrix element $\langle m|r |n \rangle$ weights the transition probability between the energy states involved.  We consider the case for which  dipole transitions that take place  between adjacent energy states are dominant. If so, 
Eq.~(\ref{equ:dipole1}) becomes
\begin{equation}
\gamma_{0} r \approx \gamma_{0} \sum_{n} \left \{ \langle n+1|r|n\rangle |n+1\rangle \langle n|+\langle n|r|n+1\rangle |n\rangle \langle n+1 | \right \}.
\end{equation}
We can recast this in the form
\begin{equation}
\gamma_{0} r   \approx   \gamma_{0} \sum_{n}  \frac{\langle n+1|r|n\rangle}{f(n+1)\sqrt{n+1}}f(n+1)\sqrt{n+1}|n+1\rangle \langle n| + h.c.
\label{equ:dipole2}
\end{equation}
Additionally, on inserting the deformed ladder operators 
\begin{eqnarray} 
\hat{A}^{\dagger}  &=&  \sum_{n}  f(n+1)\sqrt{n+1}|n+1\rangle \langle n|, \\
\hat{A}  &= & \sum_{n}  f(n+1)\sqrt{n+1}|n\rangle \langle n+1|,
\end{eqnarray}
into (\ref{equ:dipole2}) we get a more compact expression of it,
\begin{equation}
\gamma_{0}r  \approx \gamma_{0} \left(\eta(\hat{n}-1)\hat{A}^{\dagger}+\hat{A}\eta(\hat{n}-1) \right),
\label{equ:dipole3}
\end{equation}
where the operator function $\eta(\hat{n})$ is such that
\begin{equation}
\langle n+1|\eta(\hat{n}-1)|n\rangle = \frac{\langle n+1|r|n\rangle}{f(n+1)\sqrt{n+1}}.
\label{equ:elementeta}
\end{equation}
So, by making use of (17), we can write the interaction Hamiltonian (\ref{equ:intham}) as
\begin{equation}
\hat{H}_{Int} \approx \sum_{ k} \left ( \hat{b}_{ k} \kappa_{k}(\hat{n})\hat{A}^{\dagger}+\hat{A}\kappa_{ k}^{\dagger}(\hat{n})\hat{b}^{\dagger}_{ k} \right),
\label{equ:intham2}
\end{equation}
where
\begin{eqnarray}
\kappa_{ k} (\hat{n}) & = & -i\frac{(\hat{\epsilon}_{ k}\cdot \gamma_{0}\hat{e}_{r})\mathcal{E}_{k}}{\hbar} \eta(\hat{n}-1)e^{i{\bf k}\cdot {\bf r}_{0}}, \nonumber \\
& = & -i\frac{(\hat{\epsilon}_{ k}\cdot \gamma_{0}\hat{e}_{r})}{\hbar} \eta(\hat{n}-1)\sqrt{\frac{\omega_{k}}{2\hbar \epsilon_{0}V}} e^{i{\bf k}\cdot {\bf r}_{0}} 
\end{eqnarray}
can be thought of as a coupling coefficient whose value depends primarily on the level of excitation of a given oscillator via the operator function $\eta(\hat{n})$. In writing the Hamiltonian in Eq.~(\ref{equ:intham2}), the rotating-wave approximation was applied by dropping the anti resonant terms like $\hat{b}_{k}\hat{A}$ and $\hat{b}^{\dagger}_{k}\hat{A}^{\dagger}$. \\

It then follows, on substituting for $\hat{H}_{S}$, $\hat{H}_{R}$, and $H_{Int}$ from Eqs. (\ref{eq:defham1}), (\ref{equ:reservoir}) and (\ref{equ:intham2}), that the total Hamiltonian (\ref{equ:totalham}) takes the explicit form
\begin{equation}
\hat{H} \approx \frac{\hbar \Omega}{2}(\hat{A}^{\dagger}\hat{A}+\hat{A}\hat{A}^{\dagger})+ \hbar \sum_{k} \omega_{k} \hat{b}^{\dagger}_{k}\hat{b}_{k} + \sum_{ k} \left ( \hat{b}_{ k} \kappa_{ k}(\hat{n})\hat{A}^{\dagger}+\hat{A}\kappa_{ k}^{\dagger}(\hat{n})\hat{b}^{\dagger}_{ k} \right).
\label{equ:totalham2}
\end{equation}
This  Hamiltonian describing the interaction of a multi-level nonlinear system with a multi-mode field as a reservoir in the dipole and rotating-wave approximations will be the starting point for the calculations that follow. 

\section{Markovian master equation}

Now we proceed to tackle the problem of establishing an equation of motion for the description of the interaction of a given nonlinear system $S$ with its environment $R$ (the latter being considered as a radiation field reservoir). It is well-known that the dynamics of an open quantum system such as this can conveniently be described using master-equation techniques. Given that we are not interested in the evolution of the whole system-environment combination but in the influence of the environment on the system alone, our goal is to derive a master equation for the approximate evolution of such a system written in terms of its reduced density operator computed via
\begin{equation}
\hat{\rho}_{S}(t) = \Tr_{R} \{ \rho_{SR} \},
\label{equ:reduced}
\end{equation}
where $\Tr_{R}$ indicates the trace over the environment variables and $\rho_{SR}$ is the density operator of the composite system $SR$. By considering the situation described by a Hamiltonian of the form $\hat{H} = \hat{H}_{S}+\hat{H}_{R}+\hat{H}_{SR}$ (where the constituent Hamiltonians describe, respectively, the system, the environment, and the interaction), the total density operator $\rho_{SR}$ satisfies the Liouville equation of motion:
\begin{equation}
\dot{\rho}(t)_{SR}= -\frac{i}{\hbar} [\hat{H},\rho_{SR}].
\label{equ:rhototal}
\end{equation}
In order to obtain the desired approximate equation of motion just for the reduced operator $\hat{\rho}_{S}(t)$, let us suppose that the system-environment interaction takes place in the regime in which the so-called Born and Markov approximations are valid. More precisely, making these approximations implies assuming the following \cite{maximilian}: 

\begin{itemize}
\item The Born approximation: The system-environment coupling is sufficiently weak and the environment is so large compared to the system that it does not change in time significantly by the interaction, whereas the system itself does.
\item The Markov approximation: The environment correlation functions decay rapidly compared to the time scale on which the state of the system evolves, which roughly speaking amounts to saying that the system has no memory of its past. 
\end{itemize}
Application of these assumptions to Eq.~(\ref{equ:rhototal}) gives rise to the following equation of motion for the reduced density operator (\ref{equ:reduced}) (for more details see Ref. \cite{maximilian})
\begin{equation}
\dot{\widetilde{\rho}}_{S}(t)= -\frac{1}{\hbar^{2}} \int_{0}^{\infty} d\tau \Tr_{R} \left \{ \left [\widetilde{H}_{SR}(t),[\widetilde{H}_{SR}(t-\tau),\widetilde{\rho}_{S}(t)\otimes \rho_{R}]\right ] \right \},
\label{equ:masterequ1}
\end{equation}
where the tilde over the operators in this expression means that they are in the interaction picture generated by $\hat{H}_{S}+\hat{H}_{R}$, and $\rho_{R}$ represents the state of the environment (which, according to the Born approximation, is regarded as changing slowly in time). The limit of integration is taken to be infinite on the basis of the Markov approximation. \\

So, in accordance with the aforementioned approach, it turns out to be convenient to work in the interaction representation. Having established the system described by Hamiltonian (\ref{equ:totalham2}), it is found that the interaction Hamiltonian in the interaction picture generated by $\hat{H}_{S}+\hat{H}_{R}$ is given by
\begin{eqnarray}
H_{Int}(t) & = & e^{i(H_{S}+H_{R})t/\hbar}H_{Int} e^{-i(H_{S}+H_{R})t/\hbar}, \\
&  = & \hbar \left(\hat{F}^{\dagger} e^{i\Omega(\hat{n})t}\widetilde{Q}(t)+e^{-i\Omega (\hat{n})t}\hat{F}\widetilde{Q}^{\dagger}(t) \right),
\label{equ:intham3}
\end{eqnarray}
where, for simplicity, we have made the identifications:
\begin{equation}
\hat{F} = \hat{A}\eta(\hat{n}-1),
\end{equation}
\begin{equation}
\widetilde{Q}(t) = e^{i\hat{H}_{R}t/\hbar} \hat{Q} e^{-i\hat{H}_{R}t/\hbar},
\label{emu:bathvars1}
\end{equation}
\begin{equation}
\hat{Q}= \sum_{k}\kappa_{k} \hat{b}_{k}, \qquad \kappa_{k} = -i\frac{(\hat{\epsilon}_{ k}\cdot \gamma_{0}\hat{e}_{r})}{\hbar} \sqrt{\frac{\omega_{k}}{2\hbar \epsilon_{0}V}} e^{i{\bf k}\cdot {\bf r}_{p}}.
\label{emu:bathvars2}
\end{equation}
The operator function $\Omega(\hat{n})$ appearing in (\ref{equ:intham3}) is the same as in (\ref{equ:freqn}). Substitution of (\ref{equ:intham3}) into (\ref{equ:masterequ1}) allows us to arrive after a long algebra at the sought master equation:
\begin{eqnarray}
\dot{\widetilde{\rho}}_{S}(t) = & - & \hat{F}^{\dagger} e^{i\Omega(\hat{n})t} \hat{F}^{\dagger} e^{i\Omega(\hat{n})}\int_{0}^{\infty}d\tau e^{-i\Omega(\hat{n})\tau}\widetilde{\rho}(t)_{S} \Tr_{R} \{Q(\tau)Q(0)\rho_{R} \} \nonumber \\
&-& \hat{F}^{\dagger} \hat{F} \int_{0}^{\infty} d\tau e^{i\Omega(\hat{n}-1)\tau}\widetilde{\rho}_{S}(t) \Tr_{R} \{Q(\tau)Q^{\dagger}(0)\rho_{R} \} \nonumber \\
& - &\hat{F} \hat{F}^{\dagger} \int_{0}^{\infty} d\tau e^{-i\Omega(\hat{n})\tau}\widetilde{\rho}_{S}(t) \Tr_{R} \{Q^{\dagger}(\tau)Q(0)\rho_{R} \} \nonumber \\
&- & e^{-i\Omega(\hat{n})t} \hat{F} e^{-i\Omega(\hat{n})t} \hat{F} \int_{0}^{\infty} d\tau e^{i\Omega(\hat{n}-1) \tau} \widetilde{\rho}_{S}(t) \Tr_{R} \{Q^{\dagger}(\tau)Q^{\dagger}(0)\rho_{R} \} \nonumber \\
& + & \hat{F}^{\dagger} e^{i\Omega(\hat{n})t} \widetilde{\rho}_{S}(t) \hat{F}^{\dagger} e^{i\Omega(\hat{n})}\int_{0}^{\infty}d\tau e^{-i\Omega(\hat{n})\tau} \Tr_{R} \{Q(0)Q(\tau)\rho_{R} \} \nonumber \\
& + & \hat{F}^{\dagger} e^{i\Omega(\hat{n})t} \widetilde{\rho}_{S}(t)e^{-i\Omega(\hat{n})t} \hat{F} \int_{0}^{\infty} d\tau e^{i\Omega(\hat{n}-1)\tau} \Tr_{R} \{Q^{\dagger}(0)Q(\tau)\rho_{R}\} \nonumber \\
& + & e^{-i\Omega(\hat{n})t} \hat{F} \widetilde{\rho}_{S}(t) \hat{F}^{\dagger}e^{i\Omega(\hat{n})t} \int_{0}^{\infty} d\tau e^{-i\Omega(\hat{n})\tau} \Tr_{R} \{ Q(0) Q^{\dagger}(\tau) \rho_{R} \} \nonumber \\
& + & e^{-i\Omega(\hat{n})t} \hat{F} \widetilde{\rho}_{S}(t) e^{-i\Omega(\hat{n})t} \hat{F} \int_{0}^{\infty} d\tau e^{i\Omega(\hat{n}-1)\tau} \Tr_{R} \{ Q^{\dagger}(0) Q^{\dagger}(\tau) \rho_{R} \} \nonumber \\
& + & \hat{F}^{\dagger}e^{i\Omega(\hat{n})t} \int_{0}^{\infty} d\tau e^{-i\Omega(\hat{n}) \tau} \widetilde{\rho}_{S}(t) \hat{F}^{\dagger}e^{i\Omega(\hat{n})t} \Tr_{R} \{ Q(\tau) Q(0) \rho_{R} \} \nonumber \\
& + & \hat{F}^{\dagger}e^{i\Omega(\hat{n})t} \int_{0}^{\infty} d\tau e^{-i\Omega(\hat{n}) \tau} \widetilde{\rho}_{S}(t) e^{-i \Omega(\hat{n})t} \hat{F} \Tr_{R}\{Q^{\dagger}(\tau)Q(0)\rho_{R}\} \nonumber \\
& + & e^{-i\Omega(\hat{n})t} \int_{0}^{\infty} d\tau e^{i\Omega(\hat{n})\tau}\hat{F} \widetilde{\rho}_{S}(t) \hat{F}^{\dagger} e^{i\Omega(\hat{n})t} \Tr_{R} \{ Q(\tau)Q^{\dagger}(0) \rho_{R} \} \nonumber \\
& + & e^{-i\Omega(\hat{n})t} \int_{0}^{\infty} d\tau e^{i\Omega(\hat{n})\tau}\hat{F} \widetilde{\rho}_{S}(t) e^{-i\Omega(\hat{n})t} \hat{F} \Tr_{R} \{ Q^{\dagger}(\tau) Q^{\dagger}(0) \rho_{R} \} \nonumber \\
& - & \widetilde{\rho}_{S}(t) \hat{F}^{\dagger} e^{i\Omega(\hat{n})t} \hat{F}^{\dagger} e^{i\Omega(\hat{n})t} \int_{0}^{\infty} d\tau e^{-i\Omega(\hat{n}+1)\tau} \Tr_{R} \{Q(0)Q(\tau)\rho_{R} \} \nonumber \\
& - & \widetilde{\rho}_{S}(t)\hat{F}^{\dagger} \hat{F} \int_{0}^{\infty} d\tau e^{-i\Omega(\hat{n}-1)\tau} \Tr_{R} \{Q(0) Q^{\dagger}(\tau)\rho_{R} \} \nonumber \\
& - & \widetilde{\rho}_{S}(t) \hat{F} \hat{F}^{\dagger} \int_{0}^{\infty} d\tau e^{i\Omega(\hat{n})\tau} \Tr_{R} \{ Q^{\dagger}(0) Q(\tau) \rho_{R} \} \nonumber \\
& - & \widetilde{\rho}_{S}(t) e^{-i\Omega(\hat{n})t} \hat{F} e^{-i\Omega(\hat{n})t} \hat{F} \int_{0}^{\infty} d\tau e^{i\Omega(\hat{n}-2)\tau} \Tr_{R} \{ Q^{\dagger}(0) Q^{\dagger} (\tau) \rho_{R} \}. 
\label{equ:masterf1}
\end{eqnarray}
In writing the equation above, we have assumed, by virtue of the Born approximation, that the density operator of the reservoir is constant in the interaction picture: $\widetilde{\rho}_{R}(t)\approx \widetilde{\rho}_{R}(0) = \rho_{R}$. In addition, the reservoir is regarded as being in a stationary state, i.e., $\left[\rho_{R},\hat{H}_{R} \right]=0$. These considerations shape the set of correlation functions describing the fluctuations of the reservoir variables $\widetilde{Q}$ in the state $\rho_{R}$.\\

 In several circumstances, including a reservoir at thermal equilibrium, the correlation functions involving terms like $\widetilde{Q}\widetilde{Q}$ and $\widetilde{Q}^{\dagger}\widetilde{Q}^{\dagger}$ vanish, and Eq.~(\ref{equ:masterf1}) is reduced to:
\begin{eqnarray}
\dot{\widetilde{\rho}}_{S}(t) = &-&  \int_{0}^{\infty} d\tau e^{i\Omega(\hat{n}-1)\tau} \Tr_{R} \{Q(\tau)Q^{\dagger}(0)\rho_{R} \} \hat{F}^{\dagger} \hat{F} \widetilde{\rho}_{S}(t) \nonumber \\
& - & \int_{0}^{\infty} d\tau e^{-i\Omega(\hat{n})\tau} \Tr_{R} \{Q^{\dagger}(\tau)Q(0)\rho_{R} \} \hat{F} \hat{F}^{\dagger}\widetilde{\rho}_{S}(t) \nonumber \\
& + & \hat{F}^{\dagger} e^{i\Omega(\hat{n})t} \widetilde{\rho}_{S}(t)e^{-i\Omega(\hat{n})t} \hat{F} \int_{0}^{\infty} d\tau e^{i\Omega(\hat{n}-1)\tau} \Tr_{R} \{Q^{\dagger}(0)Q(\tau)\rho_{R}\} \nonumber \\
& + & e^{-i\Omega(\hat{n})t} \hat{F} \widetilde{\rho}_{S}(t) \hat{F}^{\dagger}e^{i\Omega(\hat{n})t} \int_{0}^{\infty} d\tau e^{-i\Omega(\hat{n})\tau} \Tr_{R} \{ Q(0) Q^{\dagger}(\tau) \rho_{R} \} \nonumber \\
& + &  \int_{0}^{\infty} d\tau e^{-i\Omega(\hat{n}-1) \tau} \Tr_{R}\{Q^{\dagger}(\tau)Q(0)\rho_{R}\} \hat{F}^{\dagger}e^{i\Omega(\hat{n})t} \widetilde{\rho}_{S}(t) e^{-i \Omega(\hat{n})t} \hat{F} \nonumber \\
& + & \int_{0}^{\infty} d\tau e^{i\Omega(\hat{n})\tau} \Tr_{R} \{ Q(\tau)Q^{\dagger}(0) \rho_{R} \} e^{-i\Omega(\hat{n})t} \hat{F} \widetilde{\rho}_{S}(t) \hat{F}^{\dagger} e^{i\Omega(\hat{n})t} \nonumber \\
& - & \widetilde{\rho}_{S}(t)\hat{F}^{\dagger} \hat{F} \int_{0}^{\infty} d\tau e^{-i\Omega(\hat{n}-1)\tau} \Tr_{R} \{Q(0) Q^{\dagger}(\tau)\rho_{R} \} \nonumber \\
& - & \widetilde{\rho}_{S}(t) \hat{F} \hat{F}^{\dagger} \int_{0}^{\infty} d\tau e^{i\Omega(\hat{n})\tau} \Tr_{R} \{ Q^{\dagger}(0) Q(\tau) \rho_{R} \}.
\label{equ:masterf2}
\end{eqnarray}
On transforming this equation back to the Schr\"odinger picture using $\widetilde{\rho}_{S}(t)=e^{iH_{S}t/\hbar} \rho_{S} e^{-iH_{S}t/\hbar}$, we finally obtain the Markovian master equation for the reduced density operator of a nonlinear  oscillator driven by a thermal reservoir:
\begin{eqnarray}
\dot{\rho}_{S} & = &  \frac{\Omega}{2i} \left [\hat{A}\hat{A}^{\dagger}+\hat{A}^{\dagger}\hat{A},\rho_{S} \right]-i\left[\delta_{1}(\hat{n})\hat{F}^{\dagger} \hat{F}+\delta_{2}(\hat{n})\hat{F}\hat{F}^{\dagger},\rho_{S} \right]-i\left[\delta_{3}(\hat{n}), \hat{F}^{\dagger}\rho_{S} \hat{F} \right] \nonumber \\
& - &  i\left[\delta_{4}(\hat{n}), \hat{F}\rho_{S} \hat{F}^{\dagger} \right]- \left \{ K_{1}(\hat{n})\hat{F}^{\dagger} \hat{F}+K_{2}(\hat{n})\hat{F}\hat{F}^{\dagger},\rho_{S} \right \} \nonumber \\
& + & \left \{\hat{F}^{\dagger}\rho_{S} \hat{F},K_{3}(\hat{n}) \right \} +\left \{\hat{F}\rho_{S} \hat{F}^{\dagger},K_{4}(\hat{n}) \right \},
\label{equ:masterf3}
\end{eqnarray}
where the parenthesis $[\ , \ ]$ and $\{ \ , \ \}$ denote, respectively, the commutator and anticommutator between the operators involved. Here, we have set the coefficients
\begin{eqnarray}
K_{1}(\hat{n})+i\delta_{1}(\hat{n}) & = & \int_{0}^{\infty} d\tau e^{i\Omega (\hat{n}-1)\tau} \Tr_{R} \{Q(\tau)Q^{\dagger}(0)\rho_{R} \}, \label{equ:fac1} \\
K_{2}(\hat{n})+i\delta_{2}(\hat{n}) & = & \int_{0}^{\infty} d\tau e^{-i\Omega (\hat{n})\tau} \Tr_{R} \{Q^{\dagger}(\tau)Q(0)\rho_{R} \}, \label{equ:fac2} \\
K_{3}(\hat{n})+i\delta_{3}(\hat{n}) & = & \int_{0}^{\infty} d\tau e^{i\Omega (\hat{n}-1)\tau} \Tr_{R} \{Q^{\dagger}(0)Q(\tau)\rho_{R} \}, \label{equ:fac3} \\
K_{4}(\hat{n})+i\delta_{4}(\hat{n}) & = & \int_{0}^{\infty} d\tau e^{-i\Omega (\hat{n})\tau} \Tr_{R} \{Q(0)Q^{\dagger}(\tau)\rho_{R} \}. \label{equ:fac4}
\end{eqnarray}

At first glance, Eq.~(\ref{equ:masterf3}) does not seem to bear any resemblance to the standard structure of a Lindblad-like master equation, which is, in principle, regarded as the most general Markovian master equation \cite{lindblad,maximilian}. Nevertheless, a more detailed inspection of it allows us to realize that its constitution actually generalizes the well-known Lindblad's structure, a generalization arising from both the nonlinear nature of the oscillator and from the type of coupling of it with its environment. More precisely, such a generalization manifests itself in the appearance of two sets of number-operator-dependent coefficients, namely $K_{i}$ and $\delta_{i}$. The $K_{i}$ weight those terms involving transitions between the different energy levels of the oscillator, and the $\delta_{i}$ are frequency shifts whose effect is to give rise to additional perturbing terms involved in the evolution of the off-diagonal elements of the reduced density operator; the operator quality of these coefficients does not appear for the harmonic oscillator since then $\Omega(\hat{n})$ becomes a c-number. Finally, we see that although $\hat{F}$ and $\hat{F}^{\dagger}$ are not Lindblad operators in the usual sense \cite{maximilian}, they characterize indeed the possible quantum jumps experienced by the system, and that the strength of these jumps will be determined by the the corresponding dipole matrix element associated with the specific quantum system of interest.  \\

To evaluate the above-mentioned coefficients more precisely, let the reservoir be a multimode thermal field whose density operator can be determined by using the well-known Boltzman distribution
\begin{equation}
\rho_{R} =  \frac{e^{- \hat{H}_{R}/k_{B}T} }{\Tr \{ e^{- \hat{H}_{R}/k_{B}T} \}},
\end{equation}
where $k_{B}$ is the Boltzman constant, and, as stated before, $\hat{H}_{R}$ is the reservoir Hamiltonian (\ref{equ:reservoir}), so that the last expression above becomes
\begin{equation}
\rho_{R}  = \prod_{k} e^{-\hbar \omega_{k}\hat{b}^{\dagger}_{k}\hat{b}_{k}/k_{B}T}(1-e^{-\hbar \omega_{k}/k_{B}T}).
\end{equation}
In this case, the process of tracing over the reservoir variables (\ref{emu:bathvars1}) read with (\ref{emu:bathvars2}) in each one of Eqs. (\ref{equ:fac1})-(\ref{equ:fac4}) yields
\begin{eqnarray}
\Tr_{R} \{Q(\tau)Q^{\dagger}(0)\rho_{R} \} & = & \sum_{\lambda} \int_{0}^{\infty}d^{3}k \  e^{-ikc\tau} g( {\bf k}) |\kappa({\bf k},\lambda)|^{2} [\bar{n}(kc,T)+1], \label{equ:trace1} \\
\Tr_{R} \{Q^{\dagger}(\tau)Q(0)\rho_{R} \} & = & \sum_{\lambda} \int_{0}^{\infty}d^{3}k \  e^{ikc\tau} g({\bf k}) |\kappa({\bf k},\lambda)|^{2} \bar{n}(kc,T),\label{equ:trace2}\\
\Tr_{R} \{Q^{\dagger}(0)Q(\tau)\rho_{R} \} & = & \sum_{\lambda} \int_{0}^{\infty}d^{3}k \  e^{-ikc\tau} g({\bf k}) |\kappa({\bf k},\lambda)|^{2} \bar{n}(kc,T), \label{equ:trace3} \\
\Tr_{R} \{Q(0)Q^{\dagger}(\tau)\rho_{R} \} & = & \sum_{\lambda} \int_{0}^{\infty}d^{3}k \  e^{ikc\tau} g({\bf k}) |\kappa({\bf k},\lambda)|^{2} [\bar{n}(kc,T)+1]. \label{equ:trace4}
\end{eqnarray}
Here, we have taken into account that the summation over $k$, which amounts to summing over all the wave vectors and the polarization directions, can be replaced by the integration over the same variable  in the continuum limit, i.e., 
\begin{equation}
\sum_{k} =\sum_{{\bf k},\lambda} \to \sum_{\lambda} \int_{0}^{\infty}g(k) d^{3}k,
\end{equation}
where $g(k)$ is considered to be the density of modes available per unit frequency interval in free space, so that in the polar coordinates $(k,\theta,\phi)$ \cite{scully}
\begin{equation}
g({\bf k})d^{3}k = \frac{\omega^{2}V}{8\pi^{3}c^{3}}d\omega \sin\theta d\theta d\phi
\end{equation}
represents the total number of modes in volume $V$ between the frequencies $\omega$ and $d\omega$; $\kappa_{k}$ has been rewritten as $\kappa({\bf k},\lambda)$, and $\bar{n}(\omega=kc,T)=1/(e^{\hbar \omega/k_{B}T}-1)$.

Thus, by substituting Eqs. (\ref{equ:trace1})-(\ref{equ:trace4}) into Eqs. (\ref{equ:fac1})-(\ref{equ:fac2}), respectively, together with the use of the approximation
\begin{equation}
\int_{0}^{\infty} d\tau e^{- i(\omega-\Omega_{n})\tau} \approx \pi \delta(\omega-\Omega_{n}) + i \frac{P}{\Omega_{n}-\omega},
\end{equation}
where $\Omega_{n}= \langle n|\Omega(\hat{n})|n\rangle$, and $P$ indicates the so-called Cauchy principal value, the angular integrals involved can be performed, so that the following explicit expressions for $K_{i}(\hat{n})$ and $\delta_{i}(\hat{n})$, $i=1,...,4$, are obtained:
\begin{eqnarray}
K_{1}(\hat{n}) & = & \frac{\gamma(\hat{n}-1)}{2} \left[\bar{n}(\Omega(\hat{n}-1),T)+1 \right], \\
K_{2}(\hat{n}) & = & \frac{\gamma(\hat{n})}{2} \bar{n}(\Omega(\hat{n}),T) , \\
K_{3}(\hat{n}) & = & \frac{\gamma(\hat{n}-1)}{2} \bar{n}(\Omega(\hat{n}-1),T), \\
K_{4}(\hat{n}) & = & \frac{\gamma(\hat{n})}{2} \left[\bar{n}(\Omega(\hat{n}),T)+1 \right], \\
\delta_{1}(\hat{n}) & = & \frac{\gamma(\hat{n}-1)}{2 \pi} P \int_{0}^{\infty} \frac{\omega^{3}}{\Omega^{3}(\hat{n}-1)} \frac{[\bar{n}(\omega,T)+1]}{\Omega(\hat{n}-1)-\omega}d\omega, \\
\delta_{2}(\hat{n}) & = & -\frac{\gamma(\hat{n})}{2 \pi} P \int_{0}^{\infty} \frac{\omega^{3}}{\Omega^{3}(\hat{n})} \frac{\bar{n}(\omega,T)}{\Omega(\hat{n})-\omega}d\omega, \\
\delta_{3}(\hat{n}) & = & \frac{\gamma(\hat{n}-1)}{2 \pi} P \int_{0}^{\infty} \frac{\omega^{3}}{\Omega^{3}(\hat{n}-1)} \frac{\bar{n}(\omega,T)}{\Omega(\hat{n}-1)-\omega}d\omega, \\
\delta_{4}(\hat{n}) & = & - \frac{\gamma(\hat{n})}{2 \pi} P \int_{0}^{\infty} \frac{\omega^{3}}{\Omega^{3}(\hat{n})} \frac{[\bar{n}(\omega,T)+1]}{\Omega(\hat{n})-\omega}d\omega.
\end{eqnarray}
Here,
\begin{equation}
\frac{\gamma(\hat{n})}{2}= \frac{1}{4\pi \epsilon_{0}} \frac{2 [\Omega(\hat{n})]^{3}\gamma_{0}^{2}}{3 \hbar c^{3}}
\end{equation}
can be construed as a generalized decay coefficient in the sense that its value will depend primarily upon the frequency $\Omega(\hat{n})$ associated with the level of excitation and/or the transition between two adjacent levels of the oscillator. Thus, the set of $K_{i}(\hat{n})$ coefficients describe the transition rates between anharmonic states $|n+1\rangle \to |n\rangle$ (or $|n\rangle \to |n+1\rangle $) induced by thermal photons, $\bar{n}(\Omega(\hat{n}))$, at the specific transition frequency of the energy levels involved. On the other hand, temperature-independent contribution of the $\delta_{i}(\hat{n})$ terms account for the frequency shift due to the Stark effect, whereas the temperature-dependent piece of them is the Lamb shift \cite{gardiner}.  

\section{Application: Phase space picture of a damped Morse-like oscillator}

The master-equation approach outlined above could find wide applicability to different problems of interest, provided that we make a reasonable choice of the deformation function. We put forward the idea that a given deformation function is considered to have physical significance if it enables us to reproduce the energy spectrum of the system under study through the deformed Hamiltonian (\ref{eq:defham2}). For instance, by choosing conveniently the deformation function on the basis of this criterion, it was shown in Ref. \cite{octa2} that it is possible to describe the nonlinear nature of two types of systems: the modified and the trigonometric P\"oschl-Teller potentials; the former supporting a finite number of bound states and the latter supporting an infinite number of bound states.\\

Our goal in this section is to depict the dynamics on phase space of what is called the nonlinear Morse-like oscillator \cite{recamier1} from the perspective of the temporal evolution its associated coherent states. To this end we will make use of the phase-space description based upon the Wigner quasi-distribution function, taking into account that the oscillator is viewed as an open quantum system whose evolution is governed by the master equation (\ref{equ:masterf3}). It is worth mentioning that the free evolution of such a system has already been analyzed in Refs. \cite{recamier3,recamier4,octa3,octa4} in terms of its nonlinear coherent states. It was first found in Ref. \cite{recamier1} that by choosing the deformation function
\begin{equation}
f^{2}(\hat{n}) = 1-\chi_{a} \hat{n},
\end{equation}
where $\chi_{a}$ is an anharmonicity parameter, the deformed Hamiltonian (\ref{eq:defham2}) turns out to take the form
\begin{equation}
\hat{H}_{S} = \hbar \Omega \left[ \hat{n}+\frac{1}{2}-\chi_{a}\left(\hat{n}+\frac{1}{2} \right)^{2}-\frac{\chi_{a}}{4} \right],
\label{equ:defham3}
\end{equation}
whose spectrum is in essence the same, apart from an unimportant constant term, as that of the one dimensional Morse potential \cite{landau}
\begin{equation}
E_{n} = \hbar \omega_{e}\left( n+\frac{1}{2}\right)-\frac{\hbar \omega_{e}}{2 N+1}\left(n+\frac{1}{2}\right)^{2},
\end{equation}
provided that we make the identifications $\omega_{e}=\Omega$ and $\chi_{a}=1/(2N+1)$, with $N$ being the number of bound states corresponding to the integers $0 \le n \le N-1$. 

It also turns out that for the present deformation function one gets the following commutation relations for the the set of operators $\{ \hat{n}, \hat{A}, \hat{A}^{\dagger} \}$:
\begin{equation}
\left[\hat{A},\hat{n} \right] = \hat{A}, \qquad \left[\hat{A}^{\dagger}, \hat{n} \right]=-\hat{A}^{\dagger}, \qquad \left[\hat{A}, \hat{A}^{\dagger} \right]= 1-\chi_{a}(2\hat{n}+1),
\end{equation}
where the nonlinear operators $\hat{A}$ and $\hat{A}^{\dagger}$ have the following effect upon the number operators basis $|n\rangle$:
\begin{eqnarray}
\hat{A}|n\rangle & = & \sqrt{n(1-\chi_{a}n)}|n-1\rangle, \\
\hat{A}^{\dagger} |n\rangle & = & \sqrt{(n+1)(1-\chi_{a}(n+1))}|n+1\rangle.
\end{eqnarray}
Furthermore, it was established in Ref. \cite{octa3} that provided we confine ourselves to the first $N$ bound states $|0\rangle , \ldots |N-1\rangle$ of a Morse system, the deformed operators represent, indeed, an equivalent algebraic version of the actual lowering and raising operators for the Morse wave functions. Thus, in light of the above-mentioned results, it makes sense to consider Hamiltonian (\ref{equ:defham3}) to be a suitable algebraic model to describe the discrete part of the spectrum of a Morse potential and construct their nonlinear coherent states on the basis of the f-deformed algebra. 

With regard to the construction of the coherent states associated with the Morse-like system, we have proposed at least two options to built them. In similarity to standard harmonic oscillator coherent states  (also known as Glauber coherent states \cite{glauber}) the first option is to consider the nonlinear oscillator to be an approximate eigenstate of the deformed annihilation operator, i.e., $\hat{A}|\alpha, f \rangle \approx \alpha |\alpha, f \rangle $, where $\alpha$ is a complex parameter referred to as the size of the coherent state. The number state expansion for this class of state is 
\begin{equation}
|\alpha , f\rangle \approx N_{f,\alpha} \sum_{n=0}^{N-1} \frac{\alpha^{n}}{\sqrt{n!}f(n)!}|n\rangle,
\label{equ:aocs}
\end{equation}
where $N_{f,\alpha} = (\sum_{n=0}^{N-1}\frac{|\alpha|^{2}}{n![f(n)!]^{2}})^{-1/2}$ is a normalization factor, and $f(n)!=f(0)f(1)\cdots f(n)$, with $f$ being the deformation function. Given that we are dealing with a system having a finite number $N$ of bound states, the summation in (\ref{equ:aocs}) must end at $n=N-1$, by which these states are only approximate. For the sake of brevity, they are called AOCS, which stands for {\it annihilation operator coherent states}.

On the other hand, a second option was proposed in Ref. \cite{octa3} where after applying the corresponding f-deformed displacement operator upon the ground state of the Morse-like oscillator, the coherent states thus obtained appear to be nearly well-localized on the quantum phase space from the point of view of their Wigner distribution function. Such states are defined explicitly as
\begin{equation}
|\zeta \rangle \equiv \hat{D} (\zeta (\alpha))|0\rangle \approx \frac{1}{(1+|\zeta|^{2})^{N}}\sum_{n=0}^{N-1} {2 N \choose {n}}^{1/2}\zeta^{n} |n\rangle,
\label{equ:docs}
\end{equation}
where the deformed displacement operator $\hat{D} (\zeta (\alpha)) = \exp(\alpha \hat{A}^{\dagger}-\alpha^{\ast}\hat{A})$ is thought of as a generalization of the usual displacement operator $\hat{D}(\alpha)=\exp(\alpha \hat{a}^{\dagger}-\alpha^{\ast}\hat{a})$; the problem of disentangling the exponential was overcome by making use of Lie algebraic methods \cite{puri}. Here, for a given value of $\alpha = |\alpha|e^{i\phi}$, it has also been introduced the new complex parameter $\zeta = e^{i\phi} \tan (|\alpha|\chi_{a})$. These states, referred to as {\it deformed displacement operator coherent states} (DOCS), are also approximate in same sense as that of the AOCS. \\

So, having established our Hamiltonian model and its nonlinear coherent states, let us now analyze its evolution as an open system interacting with a thermal field by means of the master equation (\ref{equ:masterf3}). In the present context, it is convenient to represent this equation in the number states basis as follows
\begin{eqnarray}
\dot{\rho}_{m,n} & = & -i \Omega_{0}(m-n)\left(1-\chi_{a}(m+n+1) \right)\rho_{m,n}\nonumber \\
& & -K_{1}(m) \eta^{2}(m-1)(1-\chi_{a} m)m\rho_{m,n} \nonumber \\
& & - K_{1}(n) \eta^{2}(n-1)(1-\chi_{a} n)n\rho_{m,n} \nonumber \\
& & -K_{2}(m)\eta^{2}(m)(1-\chi_{a}(m+1))(m+1)\rho_{m,n}\nonumber \\
& & - K_{2}(n)\eta^{2}(n)(1-\chi_{a}(n+1))(n+1)\rho_{m,n} \nonumber \\
& & + K_{3}(n)\eta(n-1)\eta(m-1)\sqrt{(1-\chi_{a} m)(1-\chi_{a} n)}\sqrt{nm}\rho_{m-1,n-1}\nonumber \\
& & + K_{3}(m)\eta(n-1)\eta(m-1)\sqrt{(1-\chi_{a} m)(1-\chi_{a} n)}\sqrt{nm}\rho_{m-1,n-1}\nonumber \\
& & + K_{4}(m) \eta(m) \eta(n) \sqrt{(1-\chi_{a}(m+1))(1-\chi_{a}(n+1))}\rho_{m+1,n+1} \nonumber \\
& & +K_{4}(n) \eta(m) \eta(n) \sqrt{(1-\chi_{a}(m+1))(1-\chi_{a}(n+1))}\rho_{m+1,n+1}, \label{equ:masterf4}
\end{eqnarray}
where $\rho_{m,n} = \langle m |\hat{\rho}|n\rangle$ and, according to Eq.~(\ref{equ:elementeta}), for this particular system it is found that 
\begin{equation}
\eta(n) = \sqrt{\frac{2 m \Omega_{0}}{\hbar}} \frac{\langle n+1|r|n\rangle}{f(n+1)\sqrt{n+1}} = \left(N+\frac{1}{2} \right) \frac{\sqrt{(N-n)(N-n-1)}}{(N-n/2)(N-n-1/2)},
\end{equation}
which is deduced from the analytic expression of the matrix elements associated to the Morse potential \cite{carvajal} 
\begin{equation}
\langle n+\alpha'|r|n\rangle = (-1)^{\alpha'+1}\beta^{-2}N_{n}N_{n+\alpha'} \left[\frac{\Gamma(k-n-\alpha')}{(k-2n-\alpha'-1)n!} \right]
\end{equation}
for $\alpha'=1$, where, in turn, $k=2N+1=1/\chi_{a}$, $\beta$ is related to the range of it, and 
\begin{equation}
N_{n} = \left[ \frac{\beta n!(k-2n-1)}{\Gamma(k-n)} \right]^{1/2}.
\label{equ:normal}
\end{equation}
As already mentioned, we shall make use of the Wigner distribution function to give a pictorial representation of the damped dynamics of the system. By definition, such a function is given, in configuration space, by the integral \cite{kim}
\begin{eqnarray}
W(r,p;t) = \frac{1}{2\pi \hbar} \int_{-\infty}^{\infty} \rho(r+y/2,r-y/2;t)e^{-ipy/\hbar}dy,
\label{equ:wigner1}
\end{eqnarray}
where $r$ and $p$ are c-numbers, and the density matrix element $\rho(r,r';t)$ represents the state of the system at time $t$. For a Morse oscillator, let $\rho(r,r';t)$ be written as 
\begin{equation}
\rho (r,r';t) = \sum_{m,n}^{N-1}\rho_{n,m}(t) \psi_{n}(r)\psi_{m}^{\ast}(r'),
\label{equ:supfuncs}
\end{equation}
where $\psi_{n}(r)$ are the well-known Morse wave functions \cite{morse}
\begin{equation}
\psi_{n}(r) = N_{n}e^{-\xi/2}\xi^{N-n}L_{n}^{2N-2n}(\xi).
\label{equ:wavefuncs}
\end{equation}
Here, $N_{n}$ is the normalization constant given by (\ref{equ:normal}), $\xi(r)  =  (2N+1)e^{-\beta r}$ is the Morse variable, and $L_{i}^{j}$ stands for the associated Laguerre polynomials. Following the procedure outlined in Ref. \cite{frank}, where the authors studied the Wigner function of bound eigenstates of the same potential, one can obtain an explicit expresion of the sought distribution function by substituting (\ref{equ:supfuncs}), together with (\ref{equ:wavefuncs}), into (\ref{equ:wigner1}) to arrive at
\begin{eqnarray}
W(r,p;t) & = & \frac{1}{2\pi \hbar}\sum_{m,n}^{N-1}\rho_{n,m}(t)\int_{-\infty}^{\infty}e^{-ipy/\hbar}\psi_{n}(r+y/2)\psi_{m}^{\ast}(r-y/2)dy \nonumber \\
& = & \frac{2}{\pi \hbar \beta} \sum_{n,m}^{N-1}\rho_{n,m}(t) N_{n}N_{m}\xi(r)^{2N-n-m}\sum_{j=0}^{m}\sum_{k=0}^{n} {2N-m \choose {m-j}} {2N-n \choose{n-k}} \nonumber \\
& & \times \frac{(-\xi(r))^{j+k}}{j!k!}K_{j+n-(k+m)-2ip/\hbar \beta} (\xi(r)),
\end{eqnarray}
where $K_{\nu}(\xi)$ are the modified Bessel functions of the third kind, and the value of the matrix element $\rho_{n,m}(t)$ will be given by solving Eq.~(\ref{equ:masterf4}) numerically. 

In what follows we shall employ the results so far obtained to examine the behavior of the Wigner function a Morse-like oscillator possessing $N=15$ ($\chi_{a}\approx 0.032 $) bound states in terms of its coherent states. This could approximately describe  the vibrational evolution of a H$_2$ molecule in a $\Sigma$ state in the presence of a thermal resorvoir. In subsequent calculations, we have omitted the contribution of the frequency shifts $\delta_{i}$ in 
Eq.~(\ref{equ:masterf4}) since they are small enough to be disregarded for the parameters considered here.

\begin{figure}[h!]
\begin{center}
\includegraphics[width=12cm, height=17cm]{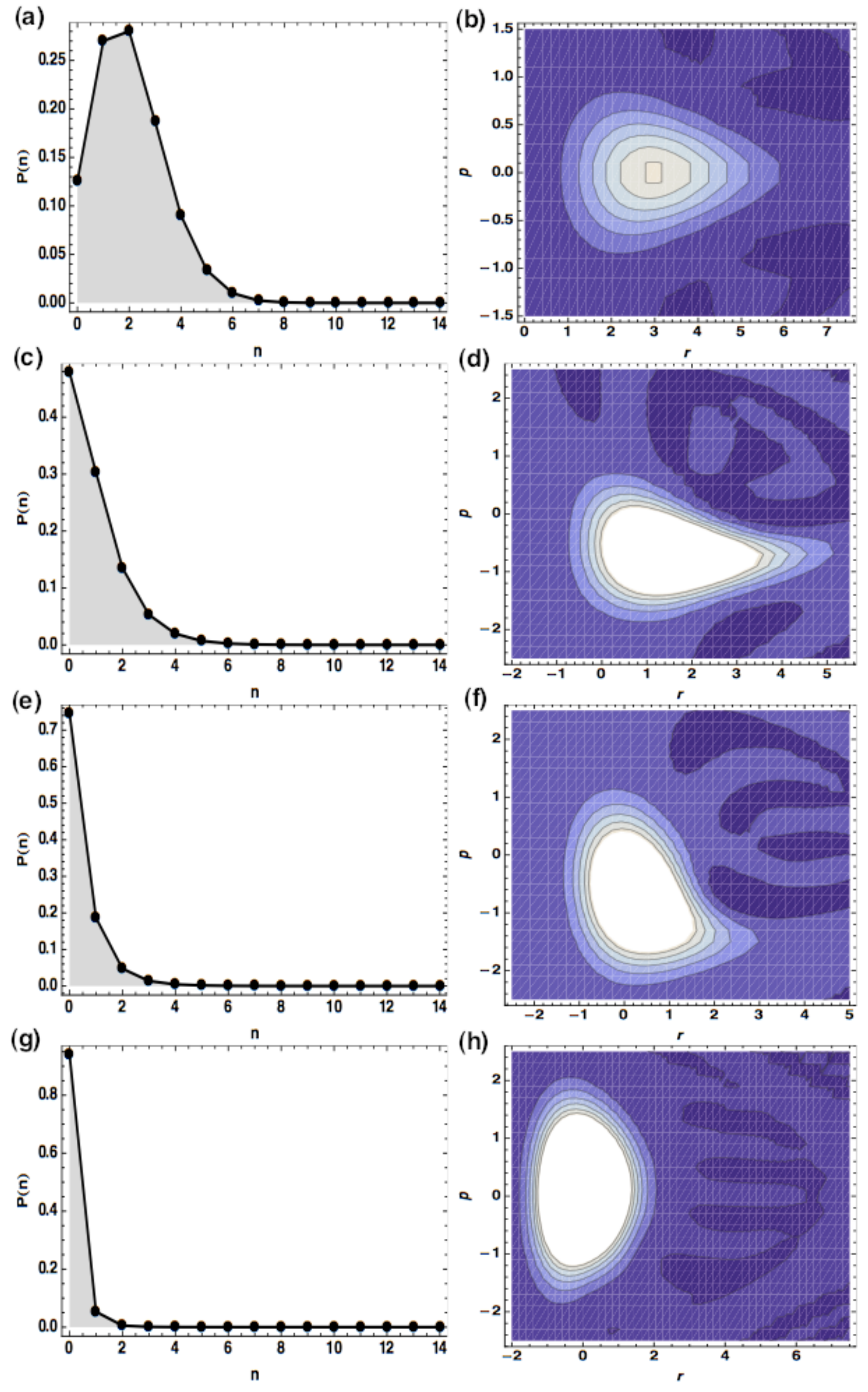} 
\end{center}
\caption{Occupation number distributions (frames (a), (c) and (e)) and the corresponding contour plots of the Wigner function (frames (b), (d) and (f)) of deformed displacement operator coherent states (DOCS) for time instants $t=0,1,2,4 \ (\Omega_{0}^{-1})$. The dimensionless displacement $r$ and momentum $p$ are depicted by the abscissa and the ordinate, respectively. As a particular case, the conditions $\hbar \Omega_{0}/k_{B}T$ =4 and $\langle \hat{n}(t=0) \rangle \approx 2$ were established in the calculation.}
\label{fig:phasespace1}
\end{figure}

As an illustrative example, we set $\hbar \Omega_{0}/k_{B}T =4$, and take $\hbar=\gamma_{0}=1$, for the sake of simplicity. 
The contour plot of the temporal evolution of the Wigner function corresponding to the DOCS from Eq.~(\ref{equ:docs}), for $\langle \hat{n}(t=0) \rangle \approx 2$ as initial condition, is displayed in Fig.~\ref{fig:phasespace1}. In the same figure it is also shown the occupation number distribution, $P(n)=|\langle n |\zeta \rangle|^{2}$, where one can see just those anharmonic phonons that play an important role in the evolution of these states. We have calculated both the phase space and the occupation number distributions for different times, say $\Omega_{0}t=0,1,2,4 $. Initially, we see from 
Fig.~\ref{fig:phasespace1} (b) that although the state of the nonlinear system is not Gaussian in form like that of the harmonic oscillator, it is somewhat well localized on phase space; it appears slightly squeezed in the $p$ direction and far more elongated in the $r$ direction. This is understandable from both the asymmetric nature of the Morse potential and the presence of potential's barrier effect, which gives rise to such a characteristic ovoid shape. Notice also that the initial Wigner function exhibits an interesting sub-Planckian structure, $i.$ $e.$, non trivial features at some regions of phase space with an area lower than $\hbar$. As time elapses, contrary to their free evolution reported in Ref.~\cite{octa3}, the Wigner function of these states does not acquire negative values, but it undergoes spreading at certain time intervals and takes positive values everywhere in the phase space, which could be interpreted as a loss of quantum coherence in their evolution due to the influence of the environment (see Fig.~\ref{fig:phasespace1} (d),(f), and (h)). This fact is compatible with the pictorial representation provided by the occupation number distribution (see Figs. (a), (c), (e), and (g)). From this point of view, one can see how the state ceases to be a DOCS and how most of its initial energy is dissipated to the environment.  

\begin{figure}[h!]
\begin{center}
\includegraphics[width=15.5cm, height=5.5cm]{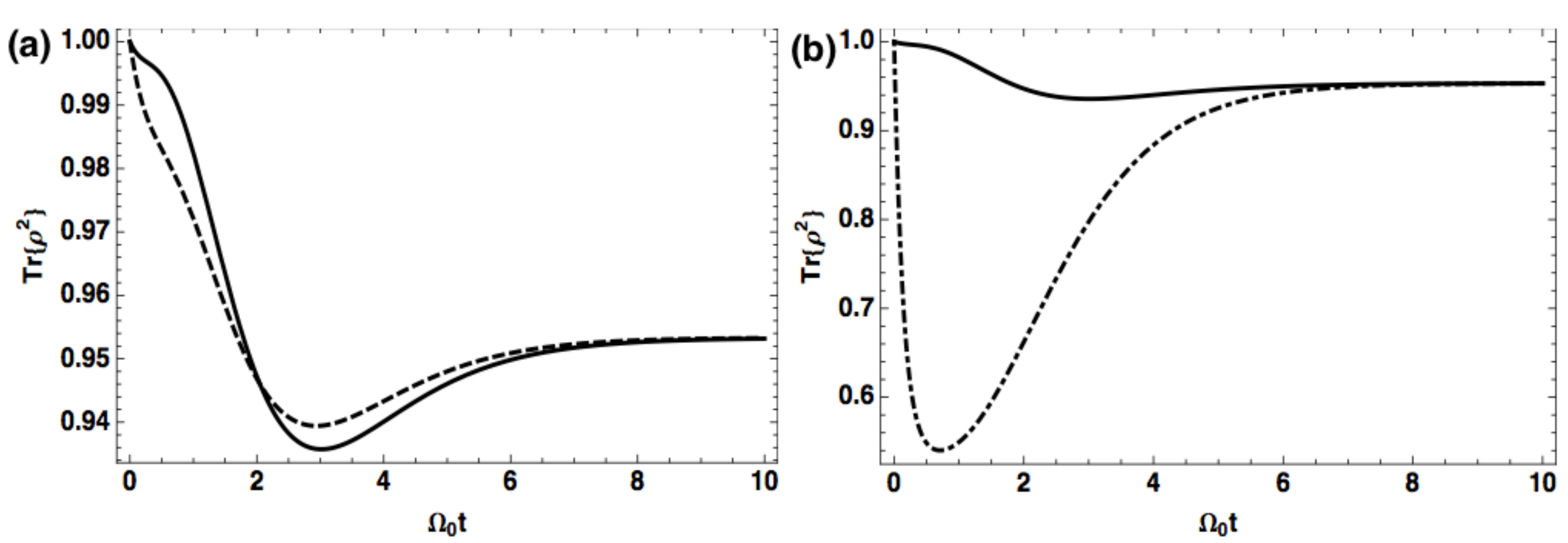}
\end{center}
\caption{(a) Evolution of the purity of a annihilation operator coherent state, AOCS, (continuos line) and a deformed displacement operator coherent state, DOCS, (dashed line) under the master equation Eq.~(\ref{equ:masterf3}). (b) Evolution of the purity of an AOCS, (continuos line) and an even superposition of two AOCS, (dot-dashed line). The initial condition $\langle \hat{n}(t=0) \rangle \approx 2$ was taken for both AOCS and DOCS, and the master equation parameters were taken as $\hbar \Omega_{0}/k_{B}T =4$, and $\gamma_{0}=1$. The unit system assumes $\hbar =1$.}
\label{fig:purity}
\end{figure}

To illustrate how the density matrix becomes a statistical mixture of number states for  nonlinear coherent states, we show in Fig. \ref{fig:purity} the evolution of the purity  defined as \begin{equation}
 Tr \rho^2 \le 1,
\end{equation}
in terms of the density matrix, or equivalently,\begin{equation}
2\pi\hbar \int W^2(r,p;t) dr dp \le 1
\end{equation}
in terms of the Wigner function, for an AOCS and a DOCS. It can be observed that the effects of decoherence due to the interaction with the thermal environment  in both DOCS and AOCS are subtle. That is, the quantum coherence of these  states is quite robust under the studied circumstances ($Tr\rho^2 \sim 1$), although  the state of the oscillator has evolved into one that cannot be considered to be close to neither a DOCS nor an AOCS.

\begin{figure}[h!]
\begin{center} 
\includegraphics[width=12cm, height=17cm]{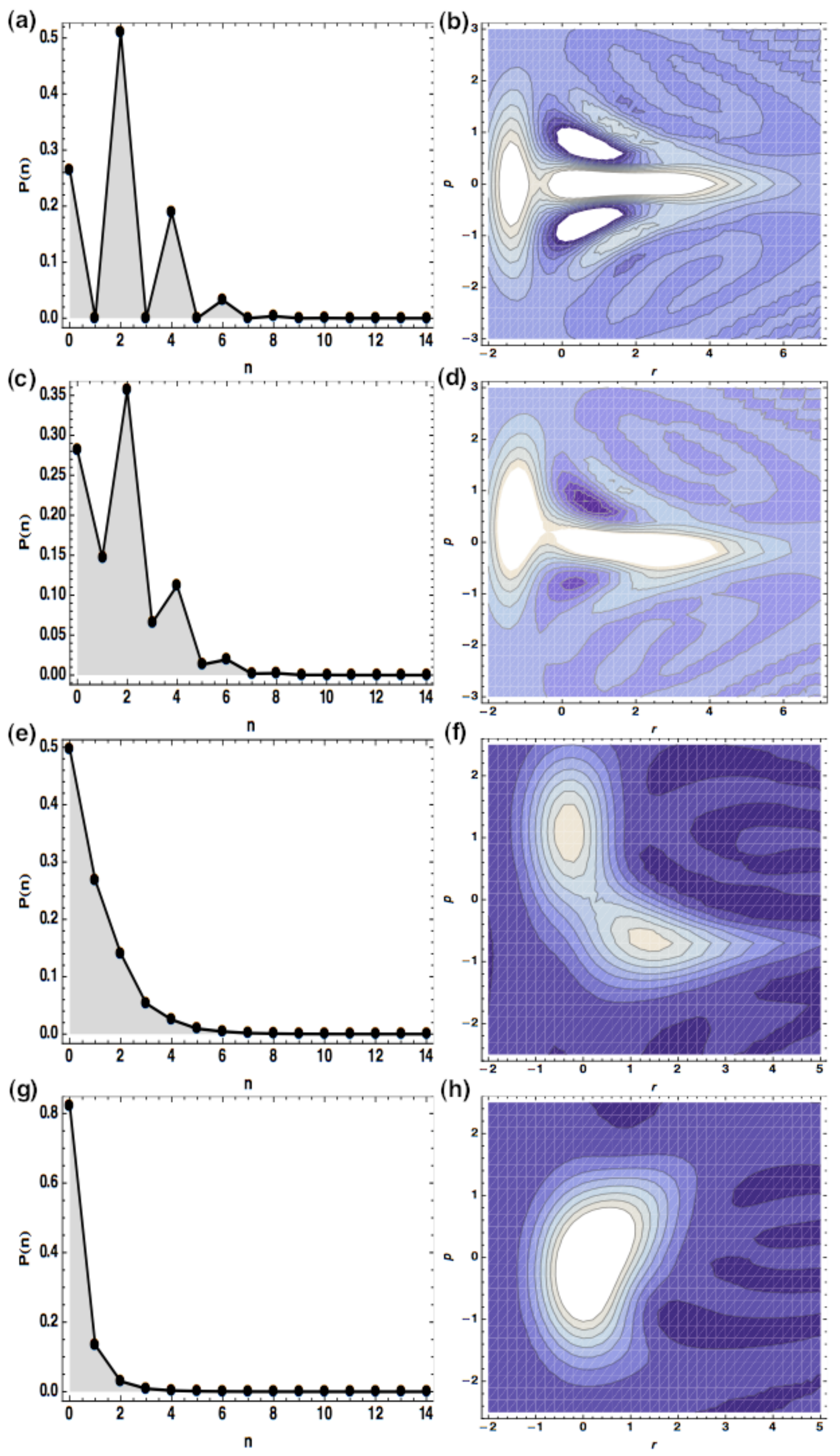} 
\end{center}
\caption{Occupation number distributions (frames (a), (c) and (e)) and the corresponding contour plots of the Wigner function (frames (b), (d) and (f)) of a superposition of two annihilation operator coherent states (AOCS) for time instants $t=0,\ 0.2,\ 1.0,\ 2.5 \ (\Omega_{0}^{-1})$. The dimensionless displacement $r$ and momentum $p$ are depicted by the abscissa and the ordinate, respectively. As a particular case, the conditions $\hbar \Omega_{0}/k_{B}T$ =4 and $\langle \hat{n}(t=0) \rangle \approx 2$ were established in the calculation.}
\label{fig:phasespace2}
\end{figure}

The decay of quantum coherence is quite more evident on superpositions of coherent states as illustrated in 
Fig.~\ref{fig:purity} (b) for the even superposition of two AOCS, i.e., a coherent state of the form $|\psi\rangle = |\alpha,f\rangle|+|-\alpha,f\rangle$. In Fig.~\ref{fig:phasespace2} we display the  corresponding evolution of the Wigner function and the occupation number distribution for the time instants $\Omega_{0}t=0,\ 0.2,\ 1.0$ and $2.5$. Here, the value of the $\alpha$ parameter is, again, such that $\langle \hat{n}(t=0) \rangle \approx 2$. Note that what started off being a coherent superposition of two states, ended up being a single incoherent hill, Fig.~\ref{fig:phasespace2} (h). Though the rough structure of a Wigner function at $\Omega_{0}t = 2.5$ is similar to that of an initial  single nonlinear coherent state at $\Omega_{0}t= 4$ Fig.~\ref{fig:phasespace1} (h), the former has lost
its quantum coherence almost completely at this stage while the purity of the latter continues being close to one. Before reaching their steady state, at $\Omega_{0}t=1$ (see Fig.~\ref{fig:phasespace2} (f)), a couple of hills representing the two components of the  superposition can be discerned; however, at this stage, of course, all the information about their initial phase relation has been lost. As expected, the dissipative properties of the interaction with the environment are reflected in the occupation number distribution showed in Figs. \ref{fig:phasespace2} (a), (c), (e), and (g). The general behavior of this property seems to be the similar to that of the single coherent state depicted in Fig.~\ref{fig:phasespace1}. The initial number state expansion of the even combination of coherent states, seen in Fig.~\ref{fig:phasespace2} (a), evolves into an incoherent mixture of number states for the first time instants (see e.g. Figs.~\ref{fig:purity} (b)). This behavior can also be understood from the fact that for sufficiently large values of $\Omega_{0}t$ only the contribution of the diagonal elements of the density matrix is preponderant, whereas the off-diagonal elements are rapidly dephased.

\section{Conclusions}

Based on the f-deformed oscillator formalism, we have introduced a generalized Born-Markov master equation for the description of the damped dynamics of nonlinear systems whose spectrum is not equally spaced. For the particular case of a thermal bath, the equation of motion thus obtained, written in terms of the deformed ladder operators, is not altogether of the Lindblad form. To illustrate its consequences, we have chosen a proper deformation function such that the corresponding deformed Hamiltonian model gives rise to a system with physical significance. This  deformation function that allows us to reproduce the energy spectrum of a Morse-like oscillator, which is in turn regarded as an open quantum system interacting with a thermal field by means of a dipolar-type interaction. From the viewpoint of the Wigner distribution function, the temporal evolution of such a system was depicted on phase space in terms of its nonlinear coherent states, defined as deformed annihilation operator coherent states (AOCS) and as deformed displacement operators coherent states (DOCS). The general behavior of dissipation and the decay of quantum coherence turned out to be essentially the same for both the AOCS and DOCS. Whereas  similar dissipation of energy into the environment but higher and faster lose of quantum coherence was observed for initial superpositions of two AOCS.

 Finally, in order to reinforce physical application of the above results, it is worth mentioning that the present approach for the said deformation function may be applied to describe the damped vibrational motion of a  diatomic  molecule modeled by a Morse potential immersed into a multimode thermal field, as long as we concentrate ourselves on the bounded part of the spectrum.

\ack This work was supported by CONACyT-M\'{e}xico through project 166961

\end{document}